\documentclass[aps,pra,10pt,twocolumn,groupedaddress,notitlepage,showpacs,floatfix]{revtex4-1}
\pdfoutput=1
\usepackage{graphicx,graphics,epsfig,subfigure,times,bm,bbm,amssymb,amsmath,amsfonts,amsthm,mathrsfs,MnSymbol}
\usepackage{gensymb}
\usepackage[matrix,frame,arrow]{xypic}
\usepackage[pdfstartview=FitH]{hyperref}
\usepackage{subfigure}
\hypersetup{
    colorlinks=true,       
    linkcolor=red,          
   citecolor=magenta,        
    filecolor=magenta,      
    urlcolor=cyan,           
    runcolor=cyan
}
\usepackage[pdftex]{color}
\usepackage{braket}
\usepackage{enumerate}
\usepackage[normalem]{ulem}
\usepackage[usenames,dvipsnames]{xcolor}
\usepackage{multirow}
\usepackage{mathtools}

\definecolor{orange}{rgb}{1,0.5,0}

\newcommand{\bes} {\begin{subequations}}
\newcommand{\ees} {\end{subequations}}
\newcommand{\bea} {\begin{eqnarray}}
\newcommand{\eea} {\end{eqnarray}}

\definecolor{gold}{rgb}{0.85,.66,0}

\newcommand{\beq}{\begin{equation}}

\newcommand{\eeq}{\end{equation}}

\newcommand{\ignore}[1]{}
\newcommand{\mc}[1]{\mathcal{#1}}

\def\tr{\mathrm{Tr}}

\def\s{\sigma}

\def\>{\rangle}
\def\<{\langle}

\def\s0{I}

\newcommand{\ig}[1]{}

\begin{document}


\title{Entanglement Swapping of Two Arbitrarily Degraded Entangled States}


\author{Brian T. Kirby}
\email[]{brian.t.kirby4.ctr@mail.mil}
\author{Siddhartha Santra}
\altaffiliation[Also at ]{Aeronautics and Astronautics, Stanford University.}
\author{Vladimir S. Malinovsky}
\author{Michael Brodsky}
\affiliation{U.S. Army Research Laboratory, Adelphi, MD 20783, USA}


\date{\today}

\begin{abstract}
 
We consider entanglement swapping, a key component of quantum network operations and entanglement distribution.
Pure entangled states, which are the desired input to the swapping protocol, are typically mixed by environmental interactions causing a reduction in their degree of entanglement.  Thus an understanding of entanglement swapping with partially mixed states is of importance.  Here we present a general analytical solution for entanglement swapping of arbitrary two-qubit states. Our result provides a comprehensive method for analyzing entanglement swapping in quantum networks. First, we show that the concurrence of a partially mixed state is conserved when this state is swapped with a Bell state. Then, we find upper and lower bounds on the concurrence of the state resulting from entanglement swapping for various classes of input states. Finally, we determine a general relationship between the ranks of the initial states and the rank of the final state after swapping. 

\end{abstract}


\maketitle

\section{Introduction \label{sec:Introduction}}
Recent interest in quantum networks is driven by the enticing possibility of powerful new network functionalities that are unattainable by conventional classical communication networks. 
Similarly to their classical counterparts, quantum networks are comprised of a multitude of nodes interconnected by quantum channels. 
While nodes generate, store and manipulate quantum states, the channels transfer or teleport these states between the nodes with high fidelity allowing the distribution of quantum entanglement across the entire network. 
This inherent ability to distribute and manipulate entanglement between distant parties is the underpinning of quantum applications.

Entanglement swapping is one of the basic quantum operations used for entanglement distribution \cite{zukowski1993event,jennewein2001experimental,de2005long}. 
For instance, it could be used for the creation of multi-partite entangled states \cite{bose1998multiparticle} from bi-partite entanglement or for overcoming the transmission loss in establishing entanglement over long-distances via quantum repeaters \cite{briegel1998quantum, duan2001long,razavi2009quantum, muralidharan2015efficient}. Interestingly, the entanglement swapping concept also lends itself to the search for entanglement conserving quantities \cite{Bose1999purification,Vogel2014unified, ge2015conservation, arkhipov2016nonclassicality}.
In any experimental implementation the generated entangled quantum states are not necessarily perfect and, in fact, could be further degraded by transmission through the communication channels. 
Entanglement swapping of certain classes of degraded states were considered recently \cite{sen2005entanglement, modlawska2008increasing, wojcik2010violation, klobus2012nonlocality, roa2014entanglement}. However, the nature of the intrinsic imperfections of any quantum network and the exact decoherence mechanism of the transmission channel itself are implementation dependent \cite{moreno2004theory, antonelli2011sudden, brodsky2011loss, shtaif2011nonlocal}. Thus the need to understand how entanglement swapping of partially degraded states works in all cases calls for a general solution for swapping of arbitrary states. 

In this paper we give an analytical description of entanglement swapping of general two-qubit states, which encompasses arbitrary two-qubit states resulting from any possible decoherence mechanisms.
While entanglement swapping can be accomplished by using projection on arbitrary basis states here we chose the Bell state basis for clarity and potential realistic implementations for photonic qubits.
We further use our analytical results to numerically model how the concurrence of the final states resulting from entanglement swapping depend on the initial states used. 
We find that entanglement swapping with any arbitrarily mixed two-qubit state represented by density matrix $M$ and a pure fully entangled Bell state results in a state, concurrence of which is the same as that of $M$. 
Next, we determine that the concurrence of a state resulting from entanglement swapping of any two Bell diagonal states is upper bounded by the product of the concurrences of the initial states. 
Lastly, we consider entanglement swapping with arbitrarily entangled pure states. We find a lower bound for this case, and discuss how the rank of input matrices affects the rank of output matrix.  Specifically we find that entanglement swapping two states of rank $R_{1}$ and $R_{2}$ results in a state with rank at least as high as $\text{max}[R_{1},R_{2}]$. Finally, our general analytical solution incorporates a few specific cases of entanglement swapping with particularly restricted input states that have been published in recent years.

The paper is organized as follows. In Sec. (\ref{sec:ent_swap}) we first obtain a closed form expression for the output two-qubit density matrix starting from two general density matrices as inputs. Numerically we build a model for optical implementation of a Bell state measurement (BSM) and use it to verify our analytical results. In Sec. (\ref{sec:Concurrence_Relations}) we present an analysis of how the concurrence of initial states is related to the concurrence of the final states using both numerical and analytical methods. We conclude with a discussion in Sec. (\ref{sec:Conclusion}).

\section{Analytical Solution for Entanglement Swapping \label{sec:ent_swap}}

\subsection{Bell states \label{subsec:E_Illustration}}


In this first subsection we start with an instructional example of entanglement swapping using Bell states.

\begin{figure}[tb]
\includegraphics[scale=0.5]{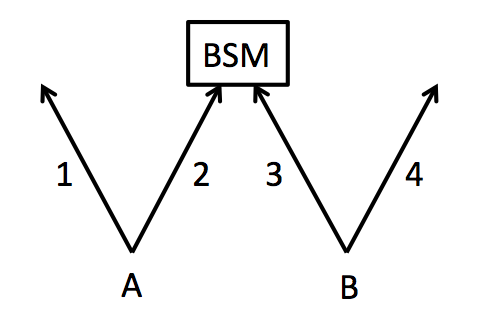}
\caption{General layout for entanglement swapping.  Source $A$ emits states entangled in modes 1 and 2, while source $B$ emits states entangled in modes 3 and 4.  Entanglement between modes 1 and 4 can sometimes be created when a BSM on modes 2 and 3 is performed.\label{fig:setup}}
\end{figure}

The setup we consider is shown in Fig.~\ref{fig:setup}, where the sources $A,B$ are independent photon-pair sources and thus the joint 4-qubit state of the photons in modes $1,2,3,4$ is given by the tensor product of the states produced at the two sources,
\begin{align}
\rho_{1,2,3,4}=\rho_{1,2}\otimes\rho_{3,4}.
\end{align}
$\mc{H}_1,\mc{H}_2$ are the Hilbert spaces of the two-qubits whose state is described by $\rho_{1,2}\in\mc{B}(\mc{H}_{1,2})\simeq\mc{B}(\mc{H}_1\otimes\mc{H}_2)$ and similarly $\mc{H}_3,\mc{H}_4$ are the spaces for the other pair $\rho_{3,4}\in\mc{B}(\mc{H}_{3,4})\simeq\mc{B}(\mc{H}_3\otimes\mc{H}_4)$, where $\mc{B}(\mc{H}_{i})$ is the space of operators on the respective Hilbert spaces.  In this subsection $\rho_{1,2}$ and $\rho_{3,4}$ are entangled two-qubit Bell states, but could represent any arbitrary two-qubit state in the rest of the paper.

If both sources $A$ and $B$ in Fig. (\ref{fig:setup}) emit the $\phi^{+}$ Bell state, then the initial system is given by:
\begin{equation}
|\phi^{+}_{1,2}\rangle \otimes |\phi^{+}_{3,4}\rangle,
\label{eq:phi_plus_initial}
\end{equation}
where the Bell states are $\vert \phi_{i,j}^{\pm}\rangle=\frac{1}{\sqrt{2}}(\vert H\rangle_{i} \vert H\rangle_{j} \pm \vert V \rangle_{i} \vert V\rangle_{j})$ and $\vert \psi_{i,j}^{\pm}\rangle=\frac{1}{\sqrt{2}}(\vert H\rangle_{i} \vert V\rangle_{j} \pm \vert V \rangle_{i} \vert H\rangle_{j})$ and $i$ and $j$ represent the modes.
As written, the state in Eq.~(\ref{eq:phi_plus_initial}) is the tensor product of modes $1,2$ and $3,4$ emitted from sources $A$ and $B$ correspondingly.  
Rewriting this composite state in terms of states with modes 2 and 3 together (the modes being measured) and modes 1 and 4 together (the modes being entangled) we obtain:
\begin{equation}
\begin{aligned}
\frac{1}{2} &[|\phi^{+}_{1,4}\rangle |\phi^{+}_{2,3}\rangle+|\phi^{-}_{1,4}\rangle |\phi^{-}_{2,3}\rangle+|\psi^{+}_{1,4}\rangle |\psi^{+}_{2,3}\rangle+|\psi^{-}_{1,4}\rangle |\psi^{-}_{2,3}\rangle].
\end{aligned}
\label{eq:swapping_bell_initial_result}
\end{equation}
Note, that a Bell State measurement is a projection of modes 2,3 onto their Bell Basis $\vert \phi^{\pm}_{2,3}\rangle$ and $\vert \psi^{\pm}_{2,3}\rangle$. 
It's clear from Eq. 3 that a BSM in modes $2,3$ will result in an entangled state in modes $1,4$, the modes which have never interacted. 
Projection of modes $2,3$ onto other entangled bases states besides the Bell basis is also capable of entangling modes $1,4$. 
However, the Bell state projection onto $\vert \psi^{-}_{2,3}\rangle$ for photonic qubits 
could be conveniently realized by using just a conventional balanced beamsplitter, hence motivating the use of BSM throughout this paper.
Information on physical implementations of BSM can be found in Subsec.~(\ref{subsec:polarization}).
In Eq.~(\ref{eq:swapping_bell_initial_result}), we see that for this particular case the final state in modes $1,4$ is the same as that found in the BSM of modes $2,3$, each outcome occurring with equal probability $\frac{1}{4}$. When swapping Bell states the output and input concurrences are all maximal and equal to 1.



\vspace{4mm}
\subsection{Arbitrary states \label{subsec:A_S_Arbitrary}}

We will now extend the above example to general density matrices in order to find the state resulting from entanglement swapping two arbitrarily mixed states. 
That is we project the joint density matrix $\rho_{1,2,3,4}$ onto a Bell state in the subspace of spatial modes $2,3$, followed by tracing these modes out and normalizing, resulting in a final state for $\rho_{14}$.  
To accomplish this projection for general input density matrices and for any Bell state outcome consider the input states for $\rho_{1,2}$ and $\rho_{3,4}$, whose elements in the basis $\vert H H \rangle, \vert H V \rangle, \vert V H \rangle, \vert V V \rangle$ are $a_{i,j}$ and $b_{i,j}$ correspondingly.
Then $\rho_{1,2,3,4}$ takes form of:
\begin{equation}
\begin{aligned}
\rho_{1,2,3,4}&=a_{11}b_{11}\vert H_{1}H_{2}\rangle \vert H_{3}H_{4} \rangle \langle H_{1}H_{2}\vert \langle H_{3}H_{4} \vert\\
&+a_{11}b_{12}\vert H_{1}H_{2}\rangle \vert H_{3}H_{4} \rangle \langle H_{1}H_{2}\vert \langle H_{3}V_{4} \vert+...
\label{eq:expanded}
\end{aligned}
\end{equation}

To facilitate the projection of modes $2$ and $3$ onto Bell states we then express all terms in $\rho_{1,2,3,4}$ as a sum of Bell states.
As an example of how this can be achieved on a term by term basis, consider the first term $\vert H_{1}H_{2}\rangle \vert H_{3}H_{4} \rangle$, which can be written as a linear sum of Bell states in modes 2 and 3 as $\frac{1}{2}(\vert \phi^{+}_{14}\rangle \vert\phi^{+}_{23}\rangle+\vert \phi^{+}_{14}\rangle \vert\phi^{-}_{23}\rangle+\vert\phi^{-}_{14}\rangle \vert\phi^{+}_{23}\rangle+\vert \phi^{-}_{14}\rangle \vert\phi^{-}_{23}\rangle)$.  
Similar results can be calculated for all 256 terms in the sum of Eq.~(\ref{eq:expanded}).  

The general output states for entanglement swapping $\rho_{1,4}^{\psi \pm}$ and $\rho_{1,4}^{\phi \pm}$ can then be found in terms of the elements of the initial input density matrices $\rho_{1,2}$ and $\rho_{3,4}$ by projecting onto either $\ket{\psi_{2,3}^{\pm}}$ or $\ket{\phi_{2,3}^{\pm}}$ respectively, followed by tracing out modes $2,3$. Thus with $\Pi^{\psi\pm}_{2,3}=\ket{{\psi^\pm}_{2,3}}\bra{{\psi^\pm}_{2,3}},\Pi^{\phi\pm}_{2,3}=\ket{{\phi^\pm}_{2,3}}\bra{{\phi^\pm}_{2,3}}$ as the projectors onto the distinct Bell states, one has: 
\begin{equation}
\rho_{1,4}^{\psi \pm}=\tr_{2,3}\bigg[\frac{\Pi^{\psi\pm}_{2,3}~\rho_{1,2,3,4}~\Pi^{\psi\pm}_{2,3}}{N_{\pm}}\bigg]
\label{eq:gen1}
\end{equation}
\begin{equation}
\rho_{1,4}^{\phi \pm}=\tr_{2,3}\bigg[\frac{\Pi^{\phi\pm}_{2,3}~\rho_{1,2,3,4}~\Pi^{\phi\pm}_{2,3}}{M_{\pm}}\bigg]
\label{eq:gen2}
\end{equation}

The output states $\rho_{1,4}^{\psi \pm}$ and $\rho_{1,4}^{\phi \pm}$ are the main analytical result of the paper and serve as the basis of our analysis presented below in Sec. \ref{sec:Concurrence_Relations} (for explicit forms of $\rho_{1,4}^{\psi \pm}$ and $\rho_{1,4}^{\phi \pm}$ and their normalization factors $N_{\pm},M_{\pm}$ see Appendix \ref{app:A}). We verified one of these results ($\rho_{1,4}^{\psi -}$) by numerical simulation for an optical implementation of a BSM that is described in Subsec. \ref{subsec:polarization}. Note that over the last few years several papers have treated entanglement swapping for particularly restricted classes of partially mixed input states, nearly all of which fall into the broad category of $X$-states \cite{wojcik2010violation, sen2005entanglement, roa2014entanglement, modlawska2008increasing, klobus2012nonlocality}. We ascertain that our general analytical solution of Eqs. (\ref{eq:gen1}) and (\ref{eq:gen2}) incorporates each of those results.  A detailed description of entanglement swapping of $X$-states is presented in Appendix \ref{subsec:E_S_x_state}.

\subsection{Photonic implementation with a beamsplitter}
\label{subsec:polarization}

Now we consider a physical implementation of the entanglement swapping protocol using polarization entangled photons and a BSM which consists of a beamsplitter and a coincidence measurement.
This model connects the results in the previous sections to realizable experiments, and also allows us to verify the results of Subsec. \ref{subsec:A_S_Arbitrary} using a formal description of the swapping setup.
As an aside, the relative simplicity of this very BSM implementation motivates the choice of the Bell basis as a basis to which project modes $2,3$.

Implementation of the BSM pictured in Fig.~\ref{fig:setup} with a $50/50$ beamsplitter and a coincidence measurement selects on the $\psi^{-}$ state by exploiting the antisymmetric nature of the singlet state that yields coincidence counts.
Identical photons will bunch at the output of a beamsplitter, which is known as the Hong-Ou-Mandel effect \cite{hong1987measurement}.
However, the opposite effect can occur if the input photons are in the singlet state, resulting in each photon exiting in a different port.
Since this anti-bunching only occurs for the singlet state we can perform a projective measurement onto $\psi_{2,3}^{-}$ by post-selecting on joint detection at the output ports of a balanced beamsplitter \cite{braunstein1995measurement, lutkenhaus1999bell, calsamiglia2001maximum, grice2011arbitrarily, berman2010principles, vogel2006quantum}.

We begin with the initial state $\rho_{1,2,3,4}=\rho_{1,2}\otimes\rho_{3,4}$ as pictured in Fig.~\ref{fig:setup}, and consider the action of a beamsplitter on modes $2,3$.
Specifically, photons in modes $2,3$ are directed as inputs to the two-input modes $a,b$ respectively of a beamsplitter (of reflectivity $\eta$) whose action on the two input modes is given by,
\begin{align}
\hat{a}^\dagger_i&\to i\sqrt{\eta}\hat{a}^\dagger_i+\sqrt{1-\eta}\hat{b}^\dagger_i\nonumber,\\
\hat{b}^\dagger_j&\to\sqrt{1-\eta}\hat{a}^\dagger_j+i\sqrt{\eta}\hat{b}^\dagger_j,
\end{align}
where $i,j=\{H,V\}$ are polarization labels for the photons in modes $a,b$, for example, $\hat{a}^\dagger_{H}$ creates a horizontally polarized photon in mode $a$ of the beamsplitter etc. This means that an input pure-state to the beamsplitter, $\ket{\psi_{\text{in}}}=\ket{i}_a\ket{j}_b=\hat{a}^\dagger_i\hat{b}^\dagger_j\ket{0}$, yields an output
\begin{align}
&\ket{\psi_{\text{out}}}=\hat{U}_{BS}\ket{\psi_{\text{in}}}\nonumber\\
&=( f(\eta)\hat{a}^\dagger_i\hat{a}^\dagger_j+(1-\eta)\hat{a}^\dagger_j\hat{b}^\dagger_i-\eta\hat{a}^\dagger_i\hat{b}^\dagger_j+ f(\eta)\hat{b}^\dagger_i\hat{b}^\dagger_j)\ket{0},
\label{psiout}
\end{align}
where, $f(\eta)=i\sqrt{\eta(1-\eta)}$. From Eq.~(\ref{psiout}) one can see that terms such as $\hat{a}^\dagger_i\hat{a}^\dagger_j~ (\hat{b}^\dagger_i\hat{b}^\dagger_j)$ create two photons in the same output mode $a~ (b)$. These doubly occupied output modes lie in the complement $\mc{H}_{\text{B}}$, of the part of the Hilbert space for the input-output modes that we are interested in - the coincidence subspace $\mc{H}_{\text{C}}$. The direct sum of these two subspaces gives us the full mode space, $\mc{H}_{\text{mode}}=\mc{H}_{\text{C}}\oplus\mc{H}_{\text{B}}$, whose spans in the $\ket{i}_a\ket{j}_b$ notation are 
\begin{align}
&\mc{H}_{\text{B}}=\text{Span}\{\ket{H}_a\ket{H}_a,\ket{V}_a\ket{V}_a,\ket{H}_a\ket{V}_a,\nonumber\\&~~~~~~~~~~~~~~~~~~~~~~~\ket{H}_b\ket{H}_b,\ket{V}_b\ket{V}_b,\ket{H}_b\ket{V}_b\}, \nonumber\\
&\mc{H}_{\text{C}}=\text{Span}\ket{H}_a\ket{H}_b,\ket{H}_a\ket{V}_b,\ket{V}_a\ket{H}_b,\nonumber\\&~~~~~~~~~~~~~~~~~~~~~~~~\ket{V}_a\ket{V}_b\}.
\end{align}
Clearly, $\text{Dim}(\mc{H}_{\text{mode}})=\text{Dim}(\mc{H}_{\text{C}})+\text{Dim}(\mc{H}_{\text{B}})=4+6=10$. 

Note that the coincidence space $\mc{H}_{\text{C}}\subset \mc{H}_{\text{mode}}$ is actually an isometric embedding of $\mc{H}_2\otimes\mc{H}_3$ into $\mc{H}_{\text{mode}}$. Denoting this isometry by the map $\hat{K}$, we have that
$\hat{K}:\mc{H}_2\otimes\mc{H}_3\mapsto\mc{H}_{\text{mode}}$, $\mc{H}_{\text{mode}}\cong_{\text{isom}}\mc{H}_2\otimes\mc{H}_3$, $\hat{K}^\dagger \hat{K}=\openone_{2,3}$. Infact, the projector $\Pi$ onto the coincidence subspace is given by $\Pi=\openone_{\text{Coin}}\oplus0_\text{Bunch}=\sum_{\alpha=1}^4\ket{\alpha_i}\bra{\alpha_i}=\sum_{i=1}^4\hat{K}\ket{i}\bra{i}\hat{K}^\dagger=\hat{K}(\sum_{i=1}^4\ket{i}\bra{i})\hat{K}^\dagger=\hat{K}\hat{K}^\dagger$, where $\ket{\alpha_i},i=1,2,3,4$ is a basis for $\mc{H}_{\text{Coin}}$ and $\ket{i},i=1,2,3,4$ is a basis for $\mc{H}_2\otimes\mc{H}_3$ with $\ket{\alpha_i}=\hat{K}\ket{i}$. Clearly $\Pi^2=\hat{K}\hat{K}^\dagger \hat{K}\hat{K}^\dagger=\hat{K}(\hat{K}^\dagger \hat{K})\hat{K}^\dagger=\hat{K}\openone \hat{K}^\dagger=\Pi$ and $\Pi\mc{H}_{\text{Coin}}=\mc{H}_{\text{Coin}},\Pi\mc{H}_{\text{Bunch}}=0.$

The unitary $\hat{U}_{BS}$, introduced in Eq.~(\ref{psiout}), acts on the entire $10$-dimensional mode space $\mc{H}_{\text{mode}}$. 
Hence before considering the action of the unitary $\hat{U}_{BS}$ on the $2,3$ part of the four-qubit input state $\rho_{1,2,3,4}$ we need a basis change operator $\hat{W}$ such that,
\begin{align}
\hat{W}:(\mc{H}_2\otimes\mc{H}_3)\otimes(\mc{H}_1\otimes\mc{H}_4)\mapsto\mc{H}_{\text{mode}}\otimes(\mc{H}_1\otimes\mc{H}_4),
\end{align}
which can be achieved by $\hat{W}=\hat{K}\otimes \openone_{1,4}$. The operator $\hat{W}$ is thus a partial isometry on the original 4-qubit space whose action is to transform,
\begin{align}
 \rho_{1,2,3,4}\mapsto\rho^{\romannumeral 1}_{1,2,3,4}=&\hat{W}\rho_{1,2,3,4}\hat{W}^\dagger\nonumber\\
 =&(\hat{K}\otimes \openone_{1,4})\rho_{1,2,3,4}(\hat{K}^\dagger\otimes \openone_{1,4}).
\end{align}
In the new basis the unitary action of the beamsplitter on $\rho^{\romannumeral 1}_{1,2,3,4}$ is given by the adjoint action of the operator $\hat{U}_{BS}$ resulting in the four-qubit density matrix $\rho^{\romannumeral 2}_{1,2,3,4}$ in the $\mc{H}_{\text{mode}}\otimes(\mc{H}_1\otimes\mc{H}_4)$ basis,
\begin{align}
&\rho^{\romannumeral 2}_{1,2,3,4}=\hat{U}_{BS}\rho^{\romannumeral 1}_{1,2,3,4}\hat{U}^\dagger_{BS}\nonumber\\
&=(\hat{U}_{BS}\otimes \openone_{1,4})(\hat{K}\otimes \openone_{1,4})\rho_{1,2,3,4}(\hat{K}^\dagger\otimes \openone_{1,4})(\hat{U}^\dagger_{BS}\otimes \openone_{1,4})\nonumber\\
&=(\hat{U}_{BS}K\otimes \openone_{1,4})\rho_{1,2,3,4}(\hat{K}^\dagger\hat{U}^\dagger_{BS}\otimes \openone_{1,4}).
\end{align}

Next we consider a coincidence measurement on the two output channels of the beamsplitter.  
Since we are interested in the four-qubit state \emph{conditioned} on the detection of a photon in each of the two output modes of the beamsplitter, we need the conditional density matrix which is actually the operator $\rho^{\romannumeral 3}_{1,2,3,4}$, obtained after normalizing the projection of $\rho^{\romannumeral 2}_{1,2,3,4}$ onto the coincidence subspace $\mc{H}_{\text{Coin}}$
\begin{align}
\rho^{\romannumeral 3}_{1,2,3,4}=\frac{\Pi \rho^{\romannumeral 2}_{1,2,3,4} \Pi}{\tr(\Pi \rho^{\romannumeral 2}_{1,2,3,4} \Pi)},
\end{align}
where the trace in the denominator is the trace over the Hilbert spaces of all 4-qubits.
One now needs to transform the basis of $\rho^{\romannumeral 3}_{1,2,3,4}$ from $\mc{H}_{\text{mode}}\otimes(\mc{H}_1\otimes\mc{H}_4)$ to $(\mc{H}_2\otimes\mc{H}_3)\otimes(\mc{H}_1\otimes\mc{H}_4)$ to give $\rho^{\romannumeral 4}_{1,2,3,4}=\hat{W}^\dagger\rho^{\romannumeral 3}_{1,2,3,4} \hat{W}$.

Finally, to yield the two qubit density matrix $\rho_{\romannumeral 4}$ we take a partial trace over the Hilbert spaces of the qubits $2,3$ in $\rho^{\romannumeral 4}_{1,2,3,4}$ resulting in,
\begin{align}
\rho^{\romannumeral 5}_{1,2,3,4}&=\tr_{2,3}(\rho^{\romannumeral 4}_{1,2,3,4})\nonumber\\
&=\frac{1}{\tr(\Pi \rho^{\romannumeral 2}_{1,2,3,4} \Pi)}\sum_{i=1}^4\braket{i|\hat{W}^\dagger\Pi\rho^{\romannumeral 2}_{1,2,3,4}\Pi \hat{W}|i}
\label{rhod}
\end{align}
where $\ket{i}$ is an orthonormal (ON) basis for $\mc{H}_2\otimes\mc{H}_3$, $\sum_{i=1}^4\ket{i}\bra{i}=\openone_{2,3}$.
Putting everything together one has that,
\begin{align}
&\rho^{\romannumeral 5}_{1,2,3,4}=\nonumber\\
&\frac{1}{\mc{N}}\sum_{i=1}^4\braket{i|(\hat{K}^\dagger\hat{U}_{BS}\hat{K}\otimes\openone_{1,4})\rho_{1,2,3,4}(\hat{K}^\dagger \hat{U}^\dagger_{BS}\hat{K}\otimes\openone_{1,4})|i}
\label{eq:rhofour}
\end{align}
With 
\begin{align}
&\mc{N}=\tr(\Pi\rho^{\romannumeral 2}_{1,2,3,4}\Pi)=\nonumber\\
&\sum_{i,j=1}^4\braket{i,j|(\hat{K}^\dagger\hat{U}_{BS}\hat{K}\otimes\openone_{1,4})\rho_{1,2,3,4}(\hat{K}^\dagger \hat{U}^\dagger_{BS}\hat{K}\otimes\openone_{1,4})|i,j},
\label{sum}
\end{align}
with $\ket{i},i=1,2,3,4$ an ON basis for $\mc{H}_2\otimes\mc{H}_3$ and $\ket{j}=1,2,3,4$ an ON basis for $\mc{H}_1\otimes\mc{H}_4$.

Note, it can be shown \cite{braunstein1995measurement} that the operator $\hat{K}^\dagger\hat{U}_{BS}\hat{K}$ appearing in Eqs. (\ref{eq:rhofour}) and (\ref{sum}) may be expressed as the sum of a projector onto the $\ket{\psi^-_{2,3}}$ state of spatial modes $2,3$ and an operator that has states with one photon per spatial mode in its kernel, i.e.,
\begin{align}
\hat{K}^\dagger\hat{U}_{BS}\hat{K}=\ket{\psi^-_{2,3}}\bra{\psi^-_{2,3}}+\hat{\mathcal{O}}.
\end{align}
Here, the operator $\hat{\mathcal{O}}$ annihilates any joint state with one photon in each mode $2,3$ in our setup.  Since it is only the latter kind of states that we focus on in this work, Eq.~(\ref{eq:rhofour}) implies that the output state $\rho_{\romannumeral 4}$ is the reduced (and normalized) part on subsystems $1,4$ after projecting onto the maximally entangled antisymmetric pure state $\ket{\psi^-_{2,3}}$ in an operator expansion, $\rho_{1,2,3,4}=\sum_{j}\hat{o}^{1,4}_j\otimes\hat{o}^{2,3}_j$.  
Eq.~(\ref{eq:rhofour}) establishes the connection to the implementation free approach of Subsecs. \ref{subsec:E_Illustration}, \ref{subsec:A_S_Arbitrary} and \ref{subsec:E_S_x_state}.

The results of this subsection are derived for a typical physical setup of entanglement swapping based on photonic qubits and a BSM relying on photon anti-bunching.
For consistency, we have implemented Eq.~(\ref{eq:rhofour}) programmatically and can numerically find the final output state $\rho_{14}$ given two numerical inputs for $\rho_{12}$ and $\rho_{34}$.
By comparing these numerical outputs with those found from $\rho_{1,4}^{\psi_{-}}$ of Eq.~(\ref{eq:gen1}) for the same inputs we have concluded that the two approaches are identical.

\section{Concurrence Relations \label{sec:Concurrence_Relations}}
Using the results of the previous sections we now apply analytical and numerical methods to analyze how entanglement swaps for various types of states.
We prove some statements analytically and for others we come to conclusions based on numerical simulations with large numbers of random density matrices.

\subsection{Entanglement swapping of a general state with a Bell state \label{subsec:claim_2}}

The most important feature of Eqs. (\ref{eq:gen1}) and (\ref{eq:gen2}) is that they can accept any input density matrix.
In this subsection we make use of this in order to show that when any input density matrix $M$ of any form is swapped with a Bell state, the resulting concurrence is equal to that of $M$. In other words, the concurrence of partially mixed state is conserved when it is swapped with a Bell state.
We first illustrate this result numerically for general input density matrices by using a large number of Bures distributed random density matrices.
Then we consider the special case of $M$ being an $X$-state, and prove this claim analytically (Appendix \ref{app:analytical_B}), due to the ease at which the concurrence of an $X$-state can be calculated. In order to sample uniformly from the space of possible density matrices we have used $10^6$ random density matrices distributed according to the Bures metric \cite{bengtsson2006geometry}.
More information about how random matrices were calculated can be found in Appendix \ref{app:random}.

\begin{figure}
\includegraphics[scale=0.7]{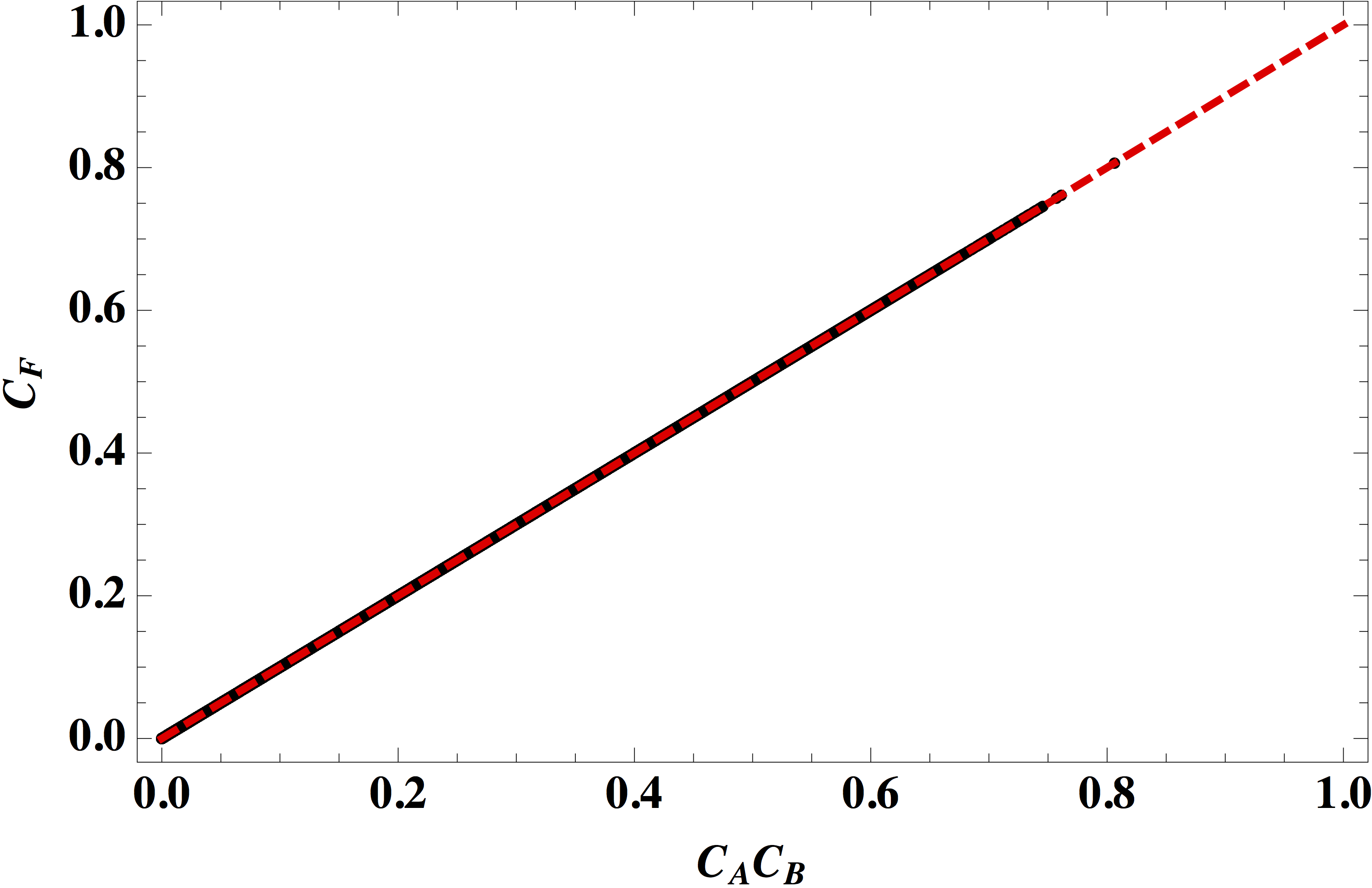}
\caption{(Color online) Resulting concurrence from entanglement swapping with a general random density matrix (with concurrence $C_{A}$) and a Bell state (with concurrence $C_{B}=1$).  The horizontal axis is the product of the initial concurrences and the vertical axis is the final concurrence ($C_{F}$) of the swapped state.  The dashed red line is the diagonal.}
\label{fig:R_vs_B}
\end{figure}

Here and in the analysis below in order to simplify the notations we label variables describing the first (second) input state in modes $1,2$ ($3,4$) as $A$ ($B$), and the final state in modes $1,4$ as $F$.

For each of the random density matrices we calculate the concurrence $C_A$, we then swap this matrix with a Bell state according to the Eqs. (\ref{eq:gen1}) and (\ref{eq:gen2}) and calculate the final concurrence $C_F$ of the resulting matrix.

Fig.~\ref{fig:R_vs_B} plots the final concurrence $C_F$ as a function of a product $C_A$ and $C_B$ ($C_B$ being a concurrence of a Bell state is equal to unity in this case) as black dots. One clearly sees that all the black dots lies exactly on the diagonal marked by dashed red line. That is the concurrence of the final state after swapping is equal to the product of the concurrences of the initial state. Note, that in this figure we project modes $2,3$ onto $\psi^{-}$ with the BSM, however the same results are found regardless of which Bell state is projected onto.

We point out that in the case considered in this section the concurrence of the initial mixed state $M$ can be viewed as a conserved quantity, because it is equal to the concurrence of the final state after entanglement swapping. Quantities which are conserved during entanglement swapping have been studied in the past \cite{Bose1999purification}, and there has been considerable recent interest in more general forms of conservation related to entanglement \cite{Vogel2014unified, ge2015conservation, arkhipov2016nonclassicality}.

This conservation result can be intuitively understood if we interpret it is as a teleportation.
Consider the case when a general state $M$ is in modes $1,2$ and the Bell state is in modes $3,4$.  
Then the BSM is implementing a teleportation of the state in mode 2 to mode 4, but without the step of applying a unitary transformation to mode $4$ to recover the original state.
However, since the entanglement of two qubit states is invariant under local unitary operations the state $\rho_{1,4}$ still has the same concurrence as the initial $M$.

\subsection{Entanglement swapping of two Bell diagonal states \label{subsec:claim_1}}

Next we consider Bell diagonal states and how the final concurrence after swapping two Bell diagonal states depends on the initial states.
We will begin this section by showing numerically that the concurrence of the final state after the entanglement swapping of two Bell diagonal states is upper bounded by the product of the concurrences of the input states.
An empirical lower bound for this case will also be found numerically in terms of the product of the input concurrences.
Lastly, we will analytically demonstrate that the swapping of the same two Bell diagonal states which are restricted to rank 2 will always swap to a state with concurrence equal to the numerical upper bound.

Bell diagonal states are a special set of $X$-states which consist of a mixture of Bell states.
For example, a general Bell diagonal state is given by:
\begin{equation}
\rho_{Bell} = \alpha \vert \psi^{+}\rangle\langle \psi^{+}\vert+\beta \vert \psi^{-}\rangle\langle \psi^{-}\vert+\gamma \vert \phi^{+}\rangle\langle \phi^{+}\vert+\delta \vert \phi^{-}\rangle\langle \phi^{-}\vert,
\label{eq:Bell_Diagonal_Definition}
\end{equation}
where the coefficients are nonnegative and sum to unity.

\begin{figure}
\includegraphics[scale=0.7]{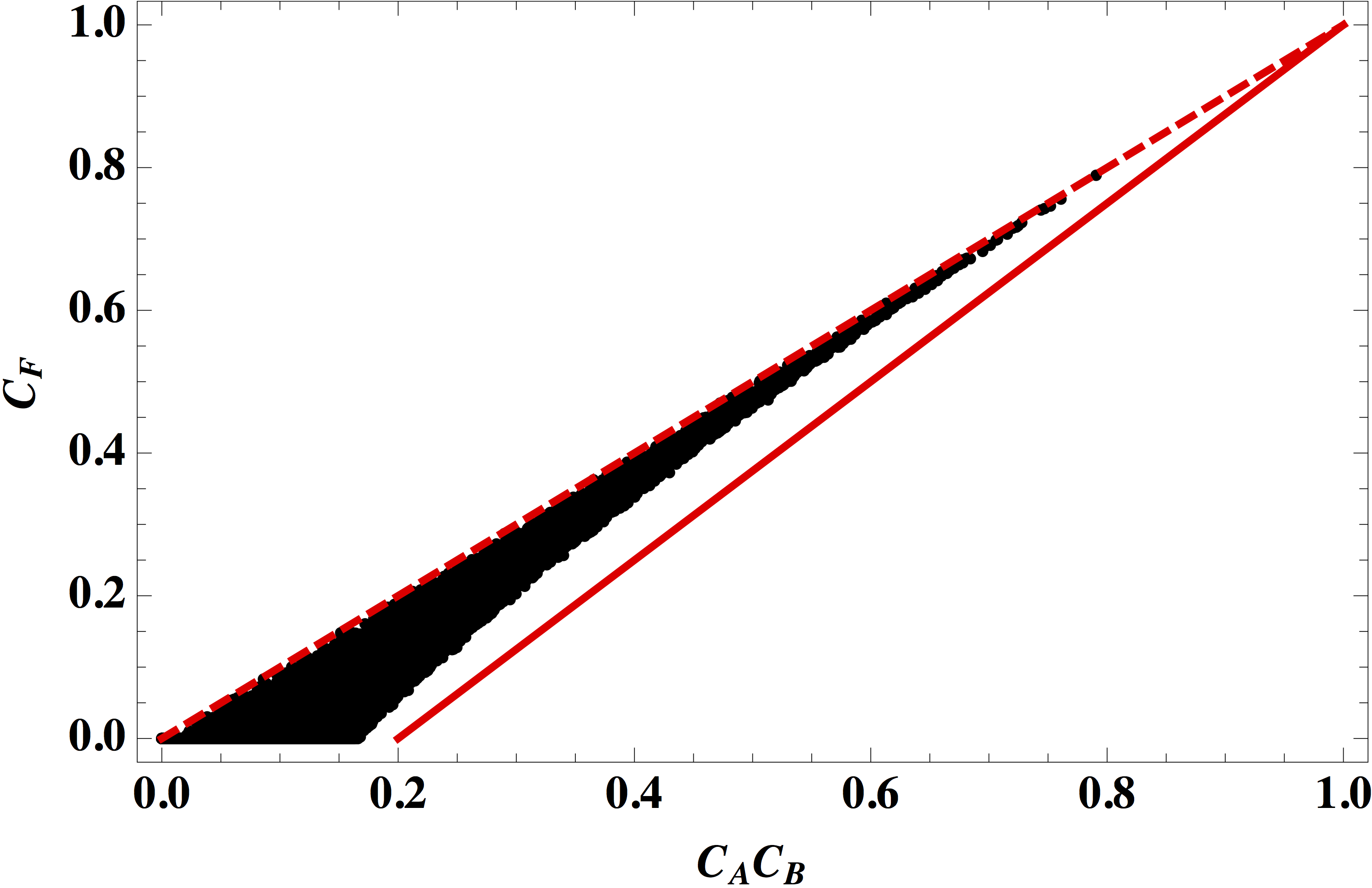}
\caption{(Color online) Entanglement swapping with $10^6$ random Bell diagonal states.  The horizontal axis is the product of the two initial concurrences (given by $C_{A}$ and $C_{B}$), and the vertical axis is the concurrence of the final state (given by $C_{F}$).  The upper and lower bounds are also displayed.\label{fig:belld}}
\end{figure}

In order to investigate how the concurrence of Bell diagonal states behaves during entanglement swapping we randomly generated $10^6$ pairs of Bell diagonal states (where the two states in the pair are in general different).
These states were generated by randomly sampling from the tetrahedron formed by Bell states \cite{bengtsson2006geometry, rocchini2000generating}.
We then swapped each of these pairs using Eqs. (\ref{eq:gen1}) and (\ref{eq:gen2}), calculated the resulting concurrence, and plotted it in Fig.~\ref{fig:belld}, where the horizontal axis is the product of the two initial states concurrences and the vertical is the final states concurrence.
This figure clearly shows there exists an upper bound on the concurrence of the final state given by the product of the input concurrences.

In addition from Fig.~\ref{fig:belld} we see that a lower bound is also present.  
For simplicity we have numerically fit this line to $\frac{5C_{A}C_{B}}{4}-\frac{1}{4}$, where $C_{A}$ and $C_{B}$ refer to the initial concurrences for the input states pictured in Fig.~\ref{fig:setup}.
These results can all be combined then to find the final inequality for entanglement swapping with Bell diagonal states: 
\begin{equation}
\text{max}\left[0,\frac{5C_{A}C_{B}}{4}-\frac{1}{4}\right]\le C_{F} \le C_{A}C_{B},
\end{equation}
where $C_{F}$ is the concurrence of the output state.

Further, the upper bound can be derived analytically for the case of swapping a rank 2 Bell diagonal state with itself.  This result is shown in Appendix \ref{app:bd_analytical}.

\subsection{Entanglement swapping of two pure states \label{subsec:claim_3}}

Several past results have indicated that swapping could result in improved final entanglement for certain input states \cite{Bose1999purification,modlawska2008increasing,wojcik2010violation,klobus2012nonlocality}.
Here we find one extremely broad class of input states (arbitrarily entangled completely pure state) is often capable of increasing the final concurrence above the product of the two initial concurrences. Moreover, the resulting state concurrence is always higher than the product of the two initial concurrences squared.

To randomly generate a large number of pure states of various degrees of entanglement we take the first column of a Haar distributed random unitary matrix \cite{mezzadri2006generate}, or equivalently, apply a Haar distributed random unitary transformation to a pure state.
More information about how we calculate Haar distributed random matrices can be found in the Appendix \ref{app:random}.

\begin{figure}
\includegraphics[scale=0.7]{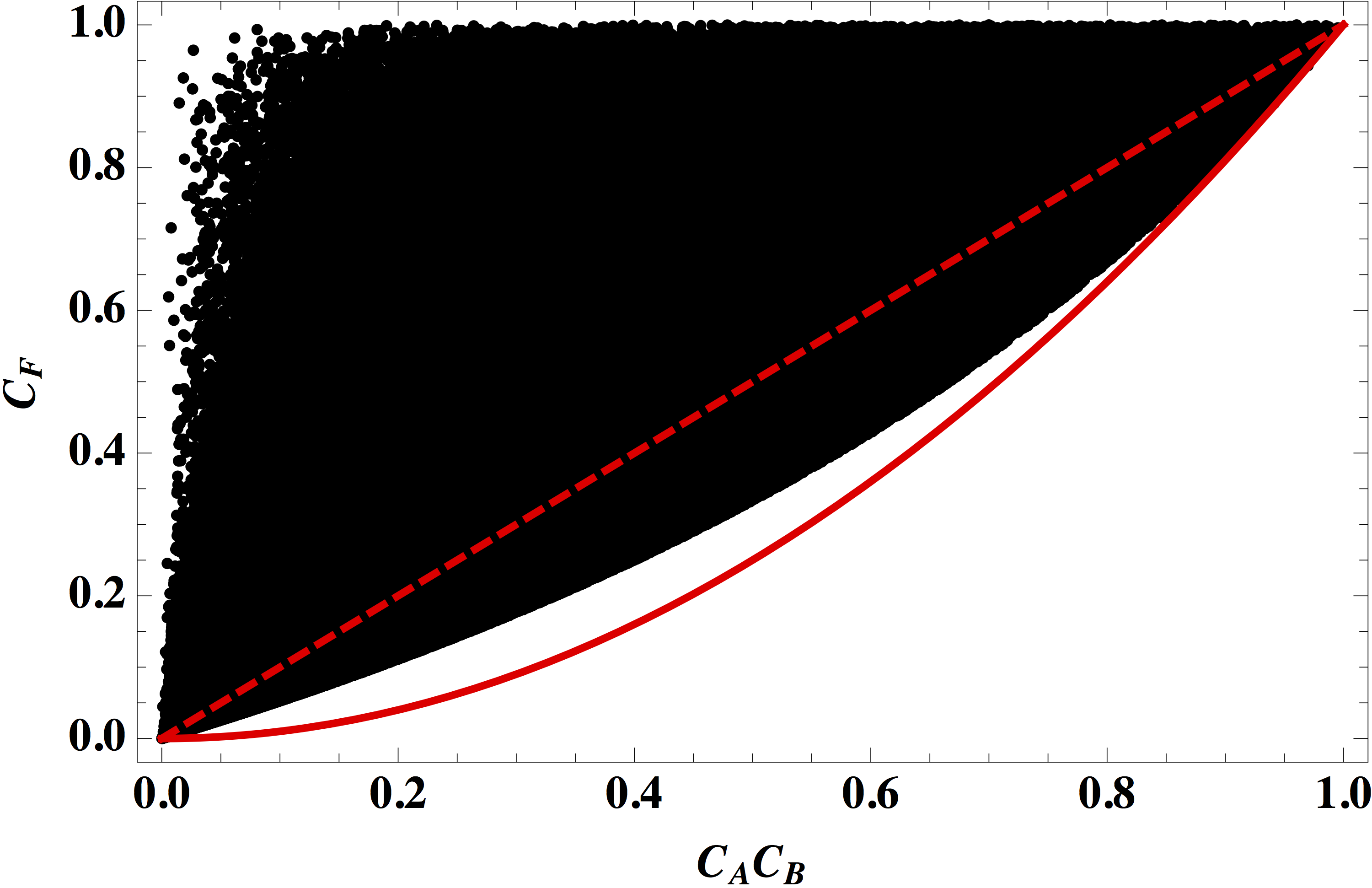}
\caption{(Color online) Comparison of concurrence before and after entanglement swapping with pure states.  The horizontal axis is the product of the two input concurrences (given by $C_{A}$ and $C_{B}$), and the vertical axis is the final concurrence (given by $C_{F}$).  The dashed line is the diagonal, and the solid line is the square of the $x$ axis.}
\label{fig:R_Pure}
\end{figure}

Pictured in Fig.~\ref{fig:R_Pure} is the result of entanglement swapping with $10^6$ pairs of random pure states, where each member of the pair is in general different.
Similarly to procedure employed for plotting previous figures, for each of the $10^6$ pairs of random matrices the final state is calculated according to Eqs.~(\ref{eq:gen1}) and (\ref{eq:gen2}); its concurrence was evaluated and plotted as a function of the product of the two initial concurrences. 
For the dataset presented in this figure we project on $\psi_{2,3}^{-}$ in the BSM, however the same results are obtained for any of the three other possible BSM outcomes.
The diagonal has also been plotted as a dashed line in Fig.~\ref{fig:R_Pure} to facilitate comparisons between the three plots.
We see that the square of the product of the initial concurrences is always lower than the final concurrence.
Although the bound is not tight it is qualitatively useful for determining the ``worst case" scenario for entanglement swapping with pure states.
For completeness we have also found an empirical lower bound from the numerical data by fitting the minimum points to an exponential given approximately by $-0.318 + 0.323 e^{1.404 C_{A}C_{B}}$, where $C_{A}$ and $C_{B}$ denote the concurrences of the two input states. It is obvious from Fig.~\ref{fig:R_Pure} that there are lots of final states whose concurrence is relatively high, and in particular is greater than the product of the two initial concurrences.

\begin{figure}
\includegraphics[scale=0.9]{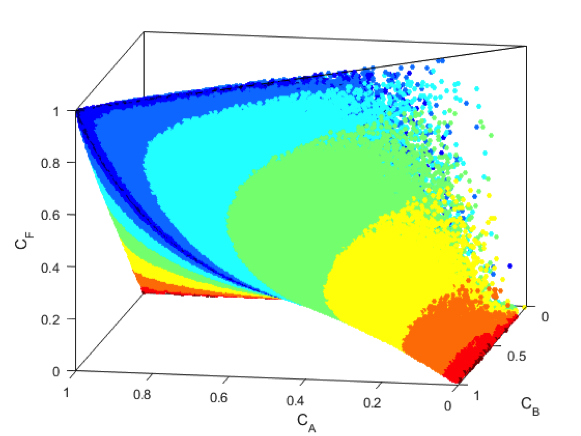}
\caption{(Color online) The concurrence ($C_{F}$) which results from swapping two randomly generated pure states of concurrence $C_{A}$ and $C_{B}$ as a function of these two concurrences.  Colored according to the ratio of the concurrences of the two input states with dark blue nearest unity.  Shown here for BSM results of $\psi^{-}$, however the density of points is identical for all BSM outcomes. }
\label{fig:colored}
\end{figure}

To clarify which pairs of states increase concurrence after swapping we consider the final concurrence as a function of both initial concurrences $C_{A}$ and $C_{B}$. Fig.~\ref{fig:colored} plots the $10^6$ points in the three spatial dimensions  $C_{A}$, $C_{B}$ and $C_{F}$. The points are colored based on the ratio of the larger of the initial concurrences to the smaller one. The color progressively changes from dark red (high ratio of the initial concurrences) to blue (ratio near unity).


The set of points in Fig.~\ref{fig:colored} appear to make up a solid bounded volume. In agreement with Fig.~\ref{fig:R_Pure} one clearly see the lower bound, but, interestingly, Fig.~\ref{fig:colored} shows an upper bound as well. The bound seems to depend on the ratio of $C_{A}C_{B}$. The main feature of the data is that the final concurrence $C_{F}$ can approach one only if the ratio of $C_{A}C_{B}$ is close to unity irrespective of the actual values of of $C_{A}$, $C_{B}$. That is the final state may be highly entangled even if the input states had low concurrences as long as those initial concurrences were near equal.

This result generalizes earlier results which show that two identical pure states which are not fully entangled can be used for purification \cite{Bose1999purification}.
In that particular case, the identical input states were imbalanced Bell states given, in the notation of section \ref{subsec:A_S_Arbitrary}, as $\cos(\theta)\vert H H \rangle + \sin(\theta)\vert V V \rangle$ with $\theta=\pi/4$ corresponding to a `balanced' input Bell state. It was found that these imbalanced states swap to the $\vert \psi^{\pm} \rangle$ Bell states upon a BSM outcome of $\vert \psi^{\pm}\rangle$.
The intuitive reason for this is that the imbalance of the input state, which is between the $\vert HH\rangle$ and $\vert VV\rangle$ terms, becomes balanced again for states with only terms involving $\vert HV \rangle$ and $\vert VH \rangle$, since each term has equal amounts of $H$ and $V$.
Since in this example the purification occurs only when the two initial states are equally imbalanced, and independent of the degree of imbalance (the angle $\theta$), we conjecture that this is the reason why Fig.~\ref{fig:colored} shows that states of similar concurrence are most likely to increase entanglement, because those are the ones which are most likely to fulfill these criteria.

\subsection{General rank relationship for any input matrices}

The above results indicate that, in general, the purity (or the rank) of a state appears to have an important impact on how it will function in an entanglement swapping setting.
To illustrate this we have considered how the rank of the output state is related to the rank of the input states.
Using $10^6$ random matrices uniformly distributed according to an induced measure (see Appendix \ref{app:random}), we have found that the rank of the final state $R_{F}$ is related to the rank of the input states $R_{A}$ and $R_{B}$ as:
\begin{equation}
R_{F}\ge\text{max}[R_{A},R_{B}].
\end{equation}
Further, we have found that the equality is satisfied when either $R_{A}$ or $R_{B}$ are equal to one.
In other words, entanglement swapping a state of rank $R$ with a pure state results in a state of the same rank $R$, and two pure states will always swap into another pure state.
Interestingly, this result is analogous to relationships between the rank of single-mode nonclassical states and their entangled two-mode outputs after a beamsplitter \cite{Vogel2014unified, sperling2011schmidt}.

\section{Conclusion \label{sec:Conclusion}}

We have given a general analytical solution for entanglement swapping of two different arbitrary bipartite states.  We have shown how this solution simplifies when input states are restricted to either $X$-states or Bell states. 
In addition, we have discussed an implementation of photonic entanglement swapping.

We have found relationships between the input and output concurrences for various classes of bipartite states. 
First, we determined that the concurrence of an arbitrary entangled state is preserved by swapping with a Bell state.
Second, through a mix of numerical and analytical means we defined both an upper and lower bound on the concurrence of a state resulting from entanglement swapping with two Bell diagonal states.  Specifically, the upper bound is the product of the two initial concurrences. 
Finally, we demonstrated the impact of purity, and rank in general, on entanglement swapping. 
We have shown that the concurrence of the final state after entanglement swapping of two pure states is lower bounded by the squared product of the concurrences of the initial states.

The reliance of future quantum networks on entanglement swapping makes these results an essential tool for predicting and understanding network performance.  Our analysis also facilitates a deeper understanding of the subtle differences of the entanglement swapping between various classes of quantum states.

\vspace{5mm}
\begin{acknowledgements}

We would like to acknowledge valuable discussions with R. Brewster, J. D. Franson, G. T. Hickman, D. E. Jones, T. B. Pittman and L. Roa.

\end{acknowledgements}

\appendix

\section{General form of output density matrices for entanglement swapping of arbitrary states}
\label{app:A}

It is convenient to use the matrix form of Eqs. (\ref{eq:gen1}) and (\ref{eq:gen2}) for calculations.  
For this reason we show here the explicit forms of the final density matrices after entanglement swapping in terms of the elements of the input density matrices:
\begin{widetext}
$\rho_{1,4}^{\psi \pm}$=
\begin{equation}
\resizebox{\linewidth}{!}{$
\frac{1}{N_{\pm}}\left(
\begin{array}{cccc}
a_{22} b_{11}\pm a_{21} b_{13}\pm a_{12} b_{31}+a_{11} b_{33} & a_{22} b_{12}\pm a_{21} b_{14}\pm a_{12} b_{32}+a_{11} b_{34} & a_{24} b_{11}\pm a_{23} b_{13}\pm a_{14} b_{31}+a_{13} b_{33} & a_{24} b_{12}\pm a_{23} b_{14}\pm a_{14} b_{32}+a_{13} b_{34} \\
 a_{22} b_{21}\pm a_{21} b_{23}\pm a_{12} b_{41}+a_{11} b_{43} & a_{22} b_{22}\pm a_{21} b_{24}\pm a_{12} b_{42}+a_{11} b_{44} & a_{24} b_{21}\pm a_{23} b_{23}\pm a_{14} b_{41}+a_{13} b_{43} & a_{24} b_{22}\pm a_{23} b_{24}\pm a_{14} b_{42}+a_{13} b_{44} \\
 a_{42} b_{11}\pm a_{41} b_{13}\pm a_{32} b_{31}+a_{31} b_{33} & a_{42} b_{12}\pm a_{41} b_{14}\pm a_{32} b_{32}+a_{31} b_{34} & a_{44} b_{11}\pm a_{43} b_{13}\pm a_{34} b_{31}+a_{33} b_{33} & a_{44} b_{12}\pm a_{43} b_{14}\pm a_{34} b_{32}+a_{33} b_{34} \\
 a_{42} b_{21}\pm a_{41} b_{23}\pm a_{32} b_{41}+a_{31} b_{43} & a_{42} b_{22}\pm a_{41} b_{24}\pm a_{32} b_{42}+a_{31} b_{44} & a_{44} b_{21}\pm a_{43} b_{23}\pm a_{34} b_{41}+a_{33} b_{43} & a_{44} b_{22}\pm a_{43} b_{24}\pm a_{34} b_{42}+a_{33} b_{44} \\
\end{array}
\right)
$}
\label{eq:ag1}
\end{equation}
$\rho_{1,4}^{\phi \pm}$=
\begin{equation}
\resizebox{\linewidth}{!}{$
\frac{1}{M_{\pm}}\left(
\begin{array}{cccc}
 a_{11} b_{11}\pm a_{12} b_{13}\pm a_{21} b_{31}+a_{22} b_{33} & a_{11} b_{12}\pm a_{12} b_{14}\pm a_{21} b_{32}+a_{22} b_{34} & a_{13} b_{11}\pm a_{14} b_{13}\pm a_{23} b_{31}+a_{24} b_{33} & a_{13} b_{12}\pm a_{14} b_{14}\pm a_{23} b_{32}+a_{24} b_{34} \\
 a_{11} b_{21}\pm a_{12} b_{23}\pm a_{21} b_{41}+a_{22} b_{43} & a_{11} b_{22}\pm a_{12} b_{24}\pm a_{21} b_{42}+a_{22} b_{44} & a_{13} b_{21}\pm a_{14} b_{23}\pm a_{23} b_{41}+a_{24} b_{43} & a_{13} b_{22}\pm a_{14} b_{24}\pm a_{23} b_{42}+a_{24} b_{44} \\
 a_{31} b_{11}\pm a_{32} b_{13}\pm a_{41} b_{31}+a_{42} b_{33} & a_{31} b_{12}\pm a_{32} b_{14}\pm a_{41} b_{32}+a_{42} b_{34} & a_{33} b_{11}\pm a_{34} b_{13}\pm a_{43} b_{31}+a_{44} b_{33} & a_{33} b_{12}\pm a_{34} b_{14}\pm a_{43} b_{32}+a_{44} b_{34} \\
 a_{31} b_{21}\pm a_{32} b_{23}\pm a_{41} b_{41}+a_{42} b_{43} & a_{31} b_{22}\pm a_{32} b_{24}\pm a_{41} b_{42}+a_{42} b_{44} & a_{33} b_{21}\pm a_{34} b_{23}\pm a_{43} b_{41}+a_{44} b_{43} & a_{33} b_{22}\pm a_{34} b_{24}\pm a_{43} b_{42}+a_{44} b_{44} \\
\end{array}
\right).
$}
\label{eq:ag2}
\end{equation}
\end{widetext}
The normalization constants are given by:
\begin{equation}
\begin{aligned}
 N_{\pm}&=a_{22} b_{11}+a_{44} b_{11}\pm a_{21} b_{13}\pm a_{43} b_{13}\\
 &+a_{22} b_{22}+a_{44} b_{22}\pm a_{21} b_{24}\pm a_{43} b_{24}\\
 &\pm a_{12} b_{31}\pm a_{34} b_{31}+a_{11} b_{33}+a_{33} b_{33}\\
 &\pm a_{12} b_{42}\pm a_{34} b_{42}+a_{11} b_{44}+a_{33} b_{44},\\
M_{\pm}&= a_{11} b_{11}+a_{33} b_{11}\pm a_{12} b_{13}\pm a_{34} b_{13}\\
&+a_{11} b_{22}+a_{33} b_{22}\pm a_{12} b_{24}\pm a_{34} b_{24}\\
&\pm a_{21} b_{31}\pm a_{43} b_{31}+a_{22} b_{33}+a_{44} b_{33}\\
&\pm a_{21} b_{42}\pm a_{43} b_{42}+a_{22} b_{44}+a_{44} b_{44}.\\
\end{aligned}
\end{equation}

\section{Analytical solution for entanglement swapping with $X$-states \label{subsec:E_S_x_state}}

Naturally, the general output states of Eqs. (\ref{eq:ag1}) and (\ref{eq:ag2}) can be simplified significantly for specific inputs such as $X$-states. Due to the considerable recent interest in $X$-states, of which the Werner states and Bell diagonal states are a special case \cite{yu2007evolution}, we will consider them in more detail here.


$X$-states are two qubit density matrices with decoupled parity sectors $\{\ket{HV},\ket{VH}\}$ and $\{\ket{HH},\ket{VV}\}$. Thus in the ordered basis for two qubits $\ket{HH},\ket{HV},\ket{VH},\ket{VV}$ an $X$-state is a density matrix of the following form:
\begin{equation}\label{eq:x_state_definition}
\chi_{c}=
\left(
\begin{array}{cccc}
 c_{11} & 0 & 0 & c_{14} \\
 0 & c_{22} & c_{23} & 0 \\
 0 & c_{32} & c_{33} & 0 \\
 c_{41} & 0 & 0 & c_{44} \\
\end{array}
\right).
\end{equation}

\vspace{5mm}

If the input density matrix for modes $1,2$ is given by Eq.~(\ref{eq:x_state_definition}), and the input for modes $3,4$ is given by
\begin{equation}
\chi_{d}=
\left(
\begin{array}{cccc}
 d_{11} & 0 & 0 & d_{14} \\
 0 & d_{22} & d_{23} & 0 \\
 0 & d_{32} & d_{33} & 0 \\
 d_{41} & 0 & 0 & d_{44} \\
\end{array}
\right),
\label{eq:x_state_definition_b}
\end{equation}
then it follows from the Eqs. (\ref{eq:gen1}) and (\ref{eq:gen2}) that the resulting output density matrices after entanglement swapping is again an $X$-state in the form of:

\begin{widetext}
\begin{equation}
\chi_{1,4}^{\psi \pm}=\frac{1}{N_{\pm}^{\chi}}
\left(
\begin{array}{cccc}
 c_{22} d_{11}+c_{11} d_{33} & 0 & 0 & \pm c_{23} d_{14}\pm c_{14} d_{32} \\
 0 & c_{22} d_{22}+c_{11} d_{44} & \pm c_{23} d_{23}\pm c_{14} d_{41} & 0 \\
 0 & \pm c_{41} d_{14}\pm c_{32} d_{32} & c_{44} d_{11}+c_{33} d_{33} & 0 \\
 \pm c_{41} d_{23}\pm c_{32} d_{41} & 0 & 0 & c_{44} d_{22}+c_{33} d_{44} \\
\end{array}
\right),
\label{eq:xr_1}
\end{equation}
\begin{equation}
\chi_{1,4}^{\phi \pm}=\frac{1}{M_{\pm}^{\chi}}
\left(
\begin{array}{cccc}
 c_{11} d_{11}+c_{22} d_{33} & 0 & 0 & \pm c_{14} d_{14}\pm c_{23} d_{32} \\
 0 & c_{11} d_{22}+c_{22} d_{44} & \pm c_{14} d_{23}\pm c_{23} d_{41} & 0 \\
 0 & \pm c_{32} d_{14}\pm c_{41} d_{32} & c_{33} d_{11}+c_{44} d_{33} & 0 \\
 \pm c_{32} d_{23}\pm c_{41} d_{41} & 0 & 0 & c_{33} d_{22}+c_{44} d_{44} \\
\end{array}
\right).
\label{eq:xr_2}
\end{equation}
\end{widetext}
The normalization constants are given by $N_{\pm}^{\chi}=c_{22} d_{11}+c_{44} d_{11}+c_{22} d_{22}+c_{44} d_{22}+c_{11} d_{33}+c_{33} d_{33}+c_{11} d_{44}+c_{33} d_{44}$ and $M_{\pm}^{\chi}=c_{11} d_{11}+c_{33} d_{11}+c_{11} d_{22}+c_{33} d_{22}+c_{22} d_{33}+c_{44} d_{33}+c_{22} d_{44}+c_{44} d_{44}$.
The results of Eqs. (\ref{eq:xr_1}) and (\ref{eq:xr_2}) agree with those of Roa et. al. \cite{roa2014entanglement}, however, their results are found by projecting onto a different set of modes than ours.


We now consider Bell state inputs as a special case of $X$-states.
This allows us to reproduce results of the example given in Sec. (\ref{subsec:E_Illustration}) using the more general formalism above.
By using the $\vert \phi^{+}\rangle$ Bell state as the input states for both, $\rho_{1,2}$ and $\rho_{3,4}$, of Eqs. (\ref{eq:ag1}) and (\ref{eq:ag2}) or of (\ref{eq:xr_1}) and (\ref{eq:xr_2}) :
\begin{equation}
\rho_{1,2}= \rho_{3,4}=\vert \phi^{+} \rangle \langle \phi^{+} \vert =
\left(
\begin{array}{cccc}
 \frac{1}{2} & 0 & 0 & \frac{1}{2} \\
 0 & 0 & 0 & 0 \\
 0 & 0 & 0 & 0 \\
 \frac{1}{2} & 0 & 0 & \frac{1}{2} \\
\end{array}
\right),
\end{equation}
we obtain:
\begin{equation}
\mu_{1,4}^{\psi \pm}=\vert \psi^{\pm} \rangle \langle \psi^{\pm} \vert =
\left(
\begin{array}{cccc}
 0 & 0 & 0 & 0 \\
 0 & \frac{1}{2} & \pm\frac{1}{2} & 0 \\
 0 & \pm\frac{1}{2} & \frac{1}{2} & 0 \\
 0 & 0 & 0 & 0 \\
\end{array}
\right),
\end{equation}
\begin{equation}
\mu_{1,4}^{\phi \pm}=\vert \phi^{\pm} \rangle \langle \phi^{\pm} \vert =
\left(
\begin{array}{cccc}
 \frac{1}{2} & 0 & 0 & \pm\frac{1}{2} \\
 0 & 0 & 0 & 0 \\
 0 & 0 & 0 & 0 \\
 \pm\frac{1}{2} & 0 & 0 & \frac{1}{2} \\
\end{array}
\right),
\end{equation}
where $\mu_{1,4}^{\psi \pm}$ and $\mu_{1,4}^{\phi \pm}$ represent the final density matrix of modes $1,4$ and the superscript indicates which Bell state was projected onto in modes $2,3$.
The output matrix is one of the four Bell states in modes $1,4$, and exactly which one is determined by the specific BSM in modes $2,3$ is performed.  This is in agreement with the results we found in Eq.~(\ref{eq:swapping_bell_initial_result}). 

Generalization to the other Bell state input combination is summarized in Table~\ref{fig:bellstateswaps}, which lists 
the resulting state in modes $1,4$ assuming the result of the BSM is $\psi^{-}$.
We have chosen to show the results for projection $\psi^{-}$ since it is the most readily implementable BSM in optical experiments as further illustrated in the next subsection.

\begingroup
\everymath{\large}
\scriptsize
\begin{table}[]
\centering
\label{fig:bellstateswaps}
\begin{tabular}{l|llll}
 & $\psi^{+}$ & $\psi^{-}$ & $\phi^{+}$ & $\phi^{-}$ \\ \hline
$\psi^{+}$   & $\psi^{-}$ & $\psi^{+}$ & $\phi^{-}$ & $\phi^{+}$ \\
$\psi^{-}$   & $\psi^{+}$ & $\psi^{-}$ & $\phi^{+}$ & $\phi^{-}$ \\
$\phi^{+}$   & $\phi^{-}$ & $\phi^{+}$ & $\psi^{-}$ & $\psi^{+}$ \\
$\phi^{-}$   & $\phi^{+}$ & $\phi^{-}$ & $\psi^{+}$ & $\psi^{-}$
\end{tabular}
\caption{Output bell states for various combinations of input Bell states when the BSM of spatial modes $2,3$ results in $\psi^{-}$.  The top row and first column represent the input states in modes $1,2$ and $3,4$, and the corresponding table element represents the final state in $1,4$ after entanglement swapping.}
\end{table}
\endgroup


\section{Analytical demonstration that swapping $X$-states and Bell states preserves the concurrence of the $X$-state}
\label{app:analytical_B}

If we assume we have an initial $X$-state given by Eq.~(\ref{eq:x_state_definition}) then the result of swapping this state with, for example, the Bell state $\phi^{-}$ results in:
\begin{equation}
\sigma_{1,4}^{\psi \pm}=\left(
\begin{array}{cccc}
 x_{11} & 0 & 0 & \mp x_{14} \\
 0 & x_{22} & \mp x_{23} & 0 \\
 0 & \mp x_{32} & x_{33} & 0 \\
 \mp x_{41} & 0 & 0 & x_{44} \\
\end{array}
\right),
\end{equation}
\begin{equation}
\sigma_{1,4}^{\phi \pm} = \left(
\begin{array}{cccc}
 x_{33} & 0 & 0 & \mp x_{32} \\
 0 & x_{44} & \mp x_{41} & 0 \\
 0 & \mp x_{14} & x_{11} & 0 \\
 \mp x_{23} & 0 & 0 & x_{22} \\
\end{array}
\right),\end{equation}
where the superscript on $\sigma$ is the result of the BSM and we have changed $c$ to $x$ to avoid confusion.
As expected all of these states are $X$-states.

The concurrence of an $X$-state has a straightforward algebraic solution \cite{yu2007evolution}. 
Specifically, the concurrence of the $X$-state in Eq.~(\ref{eq:x_state_definition}) is given by
\begin{equation}
C(\chi_{c})=2\text{max}\left[0,\vert c_{14}\vert - \sqrt{c_{22}c_{33}},\vert c_{23} \vert - \sqrt{c_{11}
c_{44}}\right].
\label{eq:x_concurrence}
\end{equation}

We can easily determine, with the use of Eq.~(\ref{eq:x_concurrence}), the concurrence of the final states.
For example for $\sigma^{\phi+}$ we find a concurrence of 
\begin{equation}
C(\sigma^{\phi+})=2\text{max}\left[0,\vert x_{32} \vert - \sqrt{x_{44} x_{11}},\vert x_{41}\vert - \sqrt{x_{33}x_{22}}\right].
\end{equation}
We can see that this is equivalent to that of the initial $X$-state concurrence, given by Eq.~(\ref{eq:x_concurrence}), because $\vert x_{41}\vert=\vert x_{14}\vert$ and $\vert x_{32}\vert=\vert x_{23}\vert$, due to the Hermiticity condition on a density matrix.
The same results are found for every combination of Bell state input and choice of BSM projection.

\section{Analytical upper bound for swapping rank 2 Bell diagonal states with themselves}
\label{app:bd_analytical}

Bell diagonal states of the form shown in Eq.~(\ref{eq:Bell_Diagonal_Definition}) which have only two nonzero coefficients are rank 2 density matrices.  We will now show analytically that for any rank 2 Bell diagonal state with concurrence $C_{r}$ that entanglement swapping this state with itself results in a state with concurrence $C_{r}^2$, which is the upper bound for the general case.
To illustrate this by a specific example, consider the input state given by:
\begin{equation}
\sigma = \alpha \vert \psi^{+}\rangle\langle \psi^{+}\vert+\beta \vert \psi^{-}\rangle\langle \psi^{-}\vert,\\
\label{eq:input_two_Bell}
\end{equation}
where $\alpha+\beta=1$.
In matrix form this state becomes
\begin{equation}
\sigma=\left(
\begin{array}{cccc}
 0 & 0 & 0 & 0 \\
 0 & \frac{1}{2} & \frac{1}{2} \left(\alpha -\beta \right) & 0 \\
 0 & \frac{1}{2} \left(\alpha -\beta \right) & \frac{1}{2} & 0 \\
 0 & 0 & 0 & 0 \\
\end{array}
\right).
\label{eq:bell_d_two}
\end{equation}
The concurrence of this state can be found from Eq.~(\ref{eq:x_concurrence}), and after algebra is given by $\vert \alpha-\beta\vert$.

Using Eq.~(\ref{eq:gen1}) we find that entanglement swapping of state Eq.~(\ref{eq:bell_d_two}) with itself results in:
\begin{equation}
\left(
\begin{array}{cccc}
 0 & 0 & 0 & 0 \\
 0 & \frac{1}{2} & -\frac{1}{2} \left(\alpha -\beta \right){}^2 & 0 \\
 0 & -\frac{1}{2} \left(\alpha -\beta \right){}^2 & \frac{1}{2} & 0 \\
 0 & 0 & 0 & 0 \\
\end{array}
\right)
\end{equation}
when $\psi^{-}$ is the result of the BSM

The resulting concurrence can again be found from Eq.~(\ref{eq:x_concurrence}), and is given by $\left(\alpha-\beta \right){}^2$, which is exactly the square of the input concurrence.
Performing this same analysis with any Bell diagonal states with only two non-zero terms and for any BSM outcome has a similar outcome.

\section{Methods for creating random density matrices \label{app:random}}

Random density matrices were calculated according to \cite{miszczak2012generating}.
As mentioned in the text we have used either an induced measure such as the Hilbert-Schmidt metric the Bures metric depending on the situation.

One way to generate a random density matrix is by starting with a pure state in a higher dimension and tracing the ancillary space out to reduce the state to the desired size. 
This procedure results in density matrices distributed according to an induced probability distribution $\mu_{n,k}$, where $k$ defines the size of the ancilla space which is to be traced out, with $n=k$ resulting in the Hilbert-Schmidt ensemble \cite{miszczak2012generating, bengtsson2006geometry}. 
An $n\times n$ density matrix distributed uniformly according to $\mu_{n,k}$ can be calculated as
\begin{equation}
\frac{G(n,k)G^{\dagger}(n,k)}{\text{tr}[G(n,k)G^{\dagger}(n,k))]},
\end{equation}
where $G(n,k)$ is an $n\times k$ Ginibre matrix.
This ensemble is used in the main text when the rank of the output density matrix is important, as in Subsec.~\ref{subsec:claim_3}, because the rank of the resulting matrix is equal to $k$ when $k\le n$.

Alternatively, an $n\times n$ Bures distributed random density matrix can be calculated from
\begin{equation}
\frac{(1+U)G(n,n)G^{\dagger}(n,n)(1+U^{\dagger})}{\text{tr}[(1+U)G(n,n)G(n,n)^{\dagger}(1+U^{\dagger})]},
\end{equation}
where $U$ is an $n\times n$ Haar distributed random unitary matrix.
In order to calculate $U$ we have used the methods described in \cite{mezzadri2006generate} which involved a $QR$ decomposition.
The eigenvalues of a unitary matrix have a magnitude of $1$ and are complex.  
If a set of unitary matrices is Haar distributed then the phases of the eigenvalues will be uniformly distributed along the unit circle.
To check this we calculated the eigenvalues of $10^6$ random $4\times4$ unitary matrices and found the mean and standard deviation to be $-0.000925005$ and $1.81386$ respectively, as expected for a uniform distribution.

\bibliography{bibliography.bib}

\end{document}